\documentclass[11pt,a4paper]{article}
\pdfoutput=1
\usepackage{jheppub}
\usepackage[T1]{fontenc}

\def\beq{\begin{equation}}
\def\bea{\begin{eqnarray}}
\def\eeq{\end{equation}}
\def\eea{\end{eqnarray}}
\def\R{$\mathcal{R}$}
\def\Z{$\mathbb{Z}$}
\def\SM{Standard Model }

\title{\boldmath  Intersecting D6-branes  on the $\mathbb{Z}_{12}$-II orientifold}
\author[a]{David Bailin}
\author[a]{Alex Love}
\affiliation[a]{Department of Physics \& Astronomy, \\ University of Sussex,\\ Brighton BN1 9QH, U.K.}
\emailAdd{d.bailin@sussex.ac.uk} 
\abstract{Much work has been done by a number of authors with the aim of constructing the supersymmetric Standard Model in  type IIA intersecting-brane theories compactified on an orientifold with various $\mathbb{Z}_N$ or $\mathbb{Z}_M \times \mathbb{Z}_N$ point groups. Here we consider  the $\mathbb{Z}_{12}$  point group which has previously received comparatively little attention. We consider intersecting D6-branes that wrap 3-cycles  consisting of a 2-cycle of the 4-dimensional lattice upon which the $\mathbb{Z}_{12}$ is realised times a 1-cycle of the remaining 2-torus. 
Our discussion is restricted to the case when these 2-cycles are ``factorisable'' in the sense discussed in \S \ref{Sfact}. Although it is possible to find models with the correct supersymmetric Standard Model quark-doublet content, we have not found it possible to obtain the correct quark-singlet content.}

\arxivnumber{1310.8215}

\begin{document}
\maketitle
\flushbottom

\section{Introduction}
The use of intersecting D6-branes in Type IIA string theory offers an attractive route to constructing the \SM  in string theory \cite{Lust:2004ks,Blumenhagen:2005mu}, and indeed an attractive model having just the spectrum of the (non-supersymmetric) \SM has been obtained by Iba\~n\'ez {\it et al.} \cite{Ibanez:2001nd}. In this approach one starts with two stacks $a$, with $N_a=3$ D6-branes, and $b$ with $N_b=2$ D6-branes, each wrapping the three large spatial dimensions plus 3-cycles of the six-demensional compactified space $Y$. Open strings beginning and ending on the stack $a$ generate the gauge group $U(3) = SU(3)_{\rm colour} \times U(1)_a$, while those that begin and end on the stack $b$ generate the gauge group $U(2) =SU(2)_L \times U(1)_b$.  Thus the non-Abelian component of the \SM gauge group is immediately assured. Further, (four-dimensional) chiral fermions  in the bi-fundamental $({\bf N}_a, \bar{\bf N}_b)=({\bf 3}, \bar{\bf 2})$ representation of $U(3) \times U(2)$ appear at the multiple intersections of the two stacks. (Here the ${\bf 3}$ representation of $U(3)$ has charge $Q_a=+1$ with respect to $U(1)_a$, and the $\bar{\bf 2}$ representation of $U(2)$ has charge $Q_b=-1$ with respect to $U(1)_b$.) This is just the representation needed for the \SM quark doublet $Q_L$. However, non-supersymmetric intersecting-brane models lead to flavour-changing neutral-current (FCNC)
processes that can only be suppressed to levels consistent with the current bounds by making the string
scale rather high, of order $10^4$ TeV, which in turn leads to fine-tuning problems \cite{Abel:2003yh}. Further, in non-supersymmetric
theories, the complex structure moduli are generally unstable \cite{Blumenhagen:2001mb}. Both of these problems are avoided if instead we seek  intersecting-brane models that yield the {\em supersymmetric} Standard Model. This is the strategy that we shall pursue in this paper.

To ensure that we obtain   $\mathcal{N}=1$ supersymmetry in the four space-time dimensions, it is necessary that the compactified space $Y$ should be a Calabi-Yau 3-fold or a toroidal orbifold $\Omega=T^6/P$, where the (discrete) point group $P$ must be a subgroup of $SU(3)$ \cite{Candelas:1985en}. (We shall only consider the latter possibility.) The requirement that the point-group generator $\theta$ acts crystallographically on the lattice $\Gamma$ that defines the torus $T^6$ then restricts $P$  to be either $\mathbb{Z}_N$, with $N=3,4,6,7,8,12$, or  $\mathbb{Z}_M \times\mathbb{Z}_N$, with $N$ a multiple of $M$ and $N=2,3,4,6$ \cite{Dixon:1985jw,Dixon:1986jc}.   The first question is whether one can find stacks $a$ and $b$, as above, whose intersections yield just the three \SM quark doublets. 
However, before proceeding further it should be noted that both of these stacks are positively charged with respect to the Ramond-Ramond (RR) 7-form gauge field to which they are ``electrically'' coupled. Since $Y$ is a compact space, the  electrical flux lines associated with the RR charges must close, which can only happen if the RR charges sum to zero. This in turn requires the introduction of negative RR charge. Anti D-branes, $\bar{\rm D}$6-branes, annihilate D6-branes, and the only feasible alternative is to use the O6-planes. These are topological defects that arise when $Y$ is an orientifold, {\it i.e.} $Y= \Omega/\mathcal{R}$, where $\mathcal{R}$ is the embedding of the world-sheet parity operator in the compactified space.  This means that every stack $\kappa=a,b, ...$ has an orientifold image $\kappa'=\mathcal{R}\kappa$, and that the stack $a$ will in general intersect with both $b$ and its orientifold image $b'$. As with the intersections of $a$ with $b$, the intersections of $a$ with $b'$ also yield chiral fermions but they are now in the representation $({\bf N}_a, {\bf N}_b)=({\bf 3}, {\bf 2})$ representation of $U(3) \times U(2)$, where the ${\bf 2}$  of $U(2)$ has charge $Q_b=+1$ with respect to $U(1)_b$. Then in order to get just the $3Q_L$ quark doublets, we require that the numbers of intersections, $a \circ b$ of $a$ with $b$, and $a \circ b'$ of $a$ with $b'$, satisfy
\beq
a \circ b + a \circ b'=3 \label{ab3}
\eeq
Of course, we must also ensure that these states have weak hypercharge $Y(Q_L)=1/6$. In general, $Y$ is a linear combination
\beq
Y=\sum _{\kappa} y_{\kappa}Q_{\kappa}
\eeq
of all of the $U(1)_{\kappa}$ charges $Q_{\kappa}$. A quark doublet arising as a $({\bf 3}, \bar{\bf 2})$ representation of $U(3) \times U(2)$ has $Y({\bf 3}, \bar{\bf 2})=y_a-y_b$, whereas the alternative has $Y({\bf 3}, {\bf 2})=y_a+y_b$. If quark doublets of {\em both} types occur, then $y_a=1/6$ and $y_b=0$. However, if there is only one type then, depending upon which, all we know is that $y_a\mp y_b=1/6$.

There have been many attempts to construct  the supersymmetric Standard Model, or something like it, using a variety of orientifolds \cite{Cvetic:2001tj}-\cite{Honecker:2013kda}. None has been completely successful, but the closest approach has probably come using the $\mathbb{Z}_6'$ orientifold. The question then arises as to whether one can do better with a different orientifold. In this paper, we address that question using the $\mathbb{Z}_{12}$-II orientifold. This orbifold (and the $\mathbb{Z}_{12}$-I orbifold)  is not completely factorisable; that is, it cannot be realised on $T^2\times T^2\times T^2$. Some of the technical problems associated with such orbifolds have been discussed in \cite{Blumenhagen:2004di}. In that paper the authors determine the non-chiral  solutions of the RR tadpole cancellation conditions  when the D6-branes lie on top of the orientifold O6-planes, the whole system satisfying (twisted) sector-by-sector RR tadpole cancellation; this is more stringent than necessary, as the vanishing of RR flux just requires overall tadpole cancellation. In what follows we consider more general configurations of intersecting (fractional) D6-branes, and attempt to construct the chiral quark, lepton and Higgs spectrum of the supersymmetric Standard Model, with the strategy of imposing overall tadpole cancellation at the end to constrain any such configurations that generate the required spectrum.
\section{The $\mathbb{Z}_{12}$ orbifolds} \label{Z12orb}
The generator $\theta$ of any abelian point group $P$ may be diagonalised using  three complex coordinates $z_k \ (k=1,2,3)$ for $T^6$ such that
\beq
\theta z_k=e^{2\pi iv_k}z_k \label{thetz}
\eeq
with $0 \leq v_k<1$ and $v_1 \pm v_2 \pm v_3=0$ so that $P \subset SU(3)$. For the $\mathbb{Z}_{12}$ point group, there are two essentially different ways to ensure the $SU(3)$ holonomy:
\bea
\mathbb{Z}_{12}{\rm -I}:&&  (v_1,v_2,v_3)=\frac{1}{12}(1,-5,4) \label{z121}\\
\mathbb{Z}_{12}{\rm-II}: && (v_1,v_2,v_3)=\frac{1}{12}(1,5,-6) \label{z122}
\eea
Both of these may be realised as Coxeter orbifolds. That is to say,  $\theta$  acts on the (six-dimensional) lattice of simple roots of a Lie algebra as a (possibly generalised) Coxeter element. For the $\mathbb{Z}_{12}$-I case we may use the lattice  $SO(8) \times SU(3)$, and for $\mathbb{Z}_{12}$-II case $SO(8) \times SU(2)\times SU(2)$.  The $SO(8)$ lattice is generated by the four simple roots $\alpha_a \ (a=1,2,...,4)$ of the $SO(8)$ Lie algebra, which 
  satisfy $\alpha _a^2=2$ and $\alpha _1.\alpha _2=-1=\alpha _2.\alpha _3= \alpha_2.\alpha _4$; the other scalar products $\alpha _1.\alpha _3=0=\alpha _3.\alpha _4=\alpha _4.\alpha _1$ are all zero. 
The order 12 generalised Coxeter element is given by
\beq
C_{SO(8)^{[3]}}:= s_1s_2s_{134}
\eeq
where the Weyl reflection $s_a$  in $\alpha_a$ acts on a general vector $x$ as
\beq
s_a({ x}):=x-(x.\alpha_a)\alpha_a
\eeq
 and $s_{134}$ is the automorphism of the $SO(8)$ Dynkin diagram that cyclically permutes the outer roots $\alpha_1\rightarrow \alpha_3 \rightarrow \alpha_4 \rightarrow \alpha_1$. ($\alpha_2$ is the central root.) Then
\beq
s_{134}(x):=x-\frac{1}{2}[(x.\alpha _1)(\alpha _1- \alpha_3)+(x.\alpha _3)(\alpha _3- \alpha_4)+(x.\alpha _4)(\alpha _4- \alpha_1)]
\eeq
$C_{SO(8)^{[3]}}$ determines the action of $\theta$ on the  four basis 1-cycles $\pi _a  \ (a=1,2, ... 4)$ of the $SO(8)$ lattice:
\bea
\theta \pi _1&=&\pi _1 +\pi _2 +\pi _3 \\
\theta \pi _2&=&-\pi _1 -\pi _2  \\
\theta \pi _3&=&\pi _1 +\pi _2 +\pi _4 \\
\theta \pi _4&=&\pi _2
\eea
The $F_4$ lattice is generated by the simple roots $\beta _a \ (a=1,2, ... 4)$ of the $F_4$ Lie algebra. They satisfy $\beta_1^2=2=\beta_2^2, \ \beta_3^2=4=\beta_4^2$ and $\beta _1.\beta _2=-1, \ \beta _2.\beta _3=-2=\beta _3.\beta _4$; the other scalar products $\beta _1.\beta _3=0=  \beta _2.\beta _4=\beta _1.\beta _4$ are all zero. .
The (ordinary) Coxeter element is
\beq
C_{F_4}:=s_1s_2s_3s_4
\eeq
where the Weyl reflection is now given by
\beq
s_a({ x}):=x-2\frac{(x.\beta_a)}{(\beta_a.\beta_a)}\beta_a
\eeq
$C_{F_4}$ also acts as the generator of $\mathbb{Z}_{12}$. However, it is easy to verify that the  $SO(8)$ and $F_4$ lattices are identical. It follows that  the  orbifolds   
$F_4\times SU(3)$ for $\mathbb{Z}_{12}$-I and $F_4 \times SU(2)\times SU(2)$ for $\mathbb{Z}_{12}$-II respectively are identical to the corrsponding $SO(8)$ orbifolds, so we shall not pursue them further.
The action of $\theta$ on the remaining two basis 1-cycles, $\pi _5$ and $\pi _6$, is different for the two $\mathbb{Z}_{12}$ orbifolds.
\bea
\mathbb{Z}_{12}{\rm-I}:&& \theta\pi_5=\pi_6-\pi_5 \quad {\rm and} \quad \theta\pi_6=-\pi_5 \\
\mathbb{Z}_{12}{\rm-II}:&& \theta\pi_5=-\pi_5 \qquad {\rm and} \qquad \theta\pi_6=-\pi_6
\eea
There are six independent 2-cycles $\pi _{a,b}$ on the $SO(8)$ lattice. They are defined as $\pi _{a,b}:=\pi _a \otimes \pi _b$ with $a,b=1,2,3,4$ and $a<b$. So for both orbifolds  there are twelve independent 3-cycles $\pi _{a,b,k}:=\pi_{a,b} \otimes \pi_k$ with $k=5,6$. 

Invariant 3-cycles are constructed by evaluating the independent combinations of the form $(1+\theta +\theta ^2+ ... +\theta ^{11})\pi _{a,b,k}$. In the $\mathbb{Z}_{12}$-I case there are only {\em two} independent invariant 3-cycles
\bea
\rho _{1} &:=& (1+\theta +\theta ^2+ ... +\theta ^{11})\pi _{2,4,6}=4(\pi _{1,2,5}-\pi _{2,4,5}-\pi _{3,4,5}+\pi _{1,3,6}+\pi _{2,3,6}+\pi _{2,4,6}) \label{ro11}\\
\rho _{2}&:=&(1+\theta +\theta ^2+ ... +\theta ^{11})\pi _{3,4,6}=4(\pi _{1,3,5}+\pi _{2,3,5}+\pi _{2,4,5}-\pi _{1,2,6}-\pi _{1,3,6}-\pi _{2,3,6}+\pi _{3,4,6})  \nonumber \\
&& \label{ro21}
\eea
However, for the $\mathbb{Z}_{12}$-II case there are {\em four}:
\bea
\rho _{1} &:=& (1+\theta +\theta ^2+ ... +\theta ^{11})\pi _{2,3,5}=6(\pi _{1,4,5}+\pi _{2,3,5}+\pi _{2,4,5}) \label{ro1}\\
\rho _{2}&:=&(1+\theta +\theta ^2+ ... +\theta ^{11})\pi _{2,4,5}=6(-\pi _{1,3,5}-\pi _{2,3,5}+\pi _{2,4,5}+\pi _{3,4,5}) \\
\rho _{3} &:=&(1+\theta +\theta ^2+ ... +\theta ^{11})\pi _{2,3,6}= 6(\pi _{1,4,6}+\pi _{2,3,6}+\pi _{2,4,6}) \\
\rho _{4}&:=&(1+\theta +\theta ^2+ ... +\theta ^{11})\pi _{2,4,6}=6(-\pi _{1,3,6}-\pi _{2,3,6}+\pi _{2,4,6}+\pi _{3,4,6}) \label{ro4}
\eea
Both of these are consistent with the cohomology of these orbifolds in the untwisted sector. Because of the smaller number of independent invariant 3-cycles, the former case has the property, also posessed by the  $\mathbb{Z}_6$ orbifold, that any supersymmetric bulk 3-cycle is automatically invariant under the orientifold action $\mathcal{R}$. The action of $\mathcal{R}$ is derived for the $\mathbb{Z}_{12}$-II case in \S \ref{RII}. (The corresponding results for the $\mathbb{Z}_{12}$-I orientifold are given in the Appendix.) Then, up to an overall multiplicative factor, all supersymmetric 3-cycles  have a common bulk part, and the differing intersection numbers needed to construct the \SM must derive solely from their differing exceptional parts.
Previous experience with the the  $\mathbb{Z}_6$ orbifold \cite{Honecker:2004kb}, as opposed to the  $\mathbb{Z}_6'$ case \cite{Bailin:2006zf}, suggests that such a structure is not rich enough to permit construction of the Standard Model. In any case, as also shown in the Appendix, the 
$\mathbb{Z}_{12}$-I orbifold only has six exceptional 3-cycles, whereas there are ten in the $\mathbb{Z}_6$ case. 
Accordingly we have not studied the $\mathbb{Z}_{12}$-I case further. Henceforth we consider only the $\mathbb{Z}_{12}$-II case. 
A general  3-cycle $\pi_{\kappa}$ is specified by the eight integer wrapping numbers $n^{\kappa}_{a,b}, n^{\kappa}_3, m^{\kappa}_3$
\beq
\pi _{\kappa} := \sum _{a,b} (n_{a,b}^{\kappa} \pi _{a,b}) \otimes (n^{\kappa}_3\pi _5+m^{\kappa}_3\pi _6) \label{kappa}
\eeq
Then the invariant bulk 3-cycle constructed from this is 
\bea
\Pi_{\kappa}^{\rm bulk}&:=&2(1+\theta +\theta ^2+ ... +\theta ^{5})\pi _{\kappa} \\
&=&\sum _{p=1}^4 A^{\kappa}_{p}\rho _{p} \label{pikapbulk}
\eea
where
\bea
A^{\kappa}_{1}&=&n^{\kappa}_3a^{\kappa}_1 \\
A^{\kappa}_{2}&=&n^{\kappa}_3 a^{\kappa}_2\\
A^{\kappa}_{3}&=&m^{\kappa}_3a^{\kappa}_1 \\
A^{\kappa}_{4}&=&m^{\kappa}_3a^{\kappa}_2
\eea
with
\bea
a^{\kappa}_1:=&&-n^{\kappa}_{1,3}+n^{\kappa}_{1,4}+n^{\kappa}_{2,3} \label{a1}\\
a^{\kappa}_2:=&&n^{\kappa}_{1,2}-n^{\kappa}_{1,3}-n^{\kappa}_{1,4}+n^{\kappa}_{2,4} \label{a2}
\eea

The intersection number $\Pi_{\kappa}^{\rm bulk} \circ \Pi_{\lambda}^{\rm bulk}$ of two bulk 3-cycles is defined as 
\beq
\Pi_{\kappa}^{\rm bulk} \circ \Pi_{\lambda}^{\rm bulk}:=\frac{1}{12} \left (\sum _{k=0}^{11} \theta ^k \pi _{\kappa} \right ) \circ \left (\sum _{\ell=0}^{11} \theta ^{\ell} \pi _{\lambda} \right ) 
\eeq 
with $\pi _{\kappa}$ and $\pi _{\lambda}$ one of the basis 3-cycles $\pi _{a,b,k}$. Then 
\bea
\rho_{1} \circ \rho_{2} &=&0=\rho_{3} \circ \rho_{4} \\
\rho_{1} \circ \rho_{3}&=&6=\rho_{2} \circ \rho_{4} \\
\rho_{1} \circ \rho_{4}&=&0=\rho_{2} \circ \rho_{3}
\eea 
and for two general bulk 3-cycles of the form (\ref{kappa}) we get
\bea
\Pi ^{\rm bulk}_{\kappa} \circ \Pi ^{\rm bulk}_{\lambda}&=&6(A^{\kappa}_{1}A^{\lambda}_{3}-A^{\kappa}_{3}A^{\lambda}_{1}+ A^{\kappa}_{2}A^{\lambda}_{4}-A^{\kappa}_{4}A^{\lambda}_{2}) \\
&=& 6(a_1^{\kappa}a_1^{\lambda}+a_2^{\kappa}a_2^{\lambda})(n_3^{\kappa}m_3^{\lambda}-m_3^{\kappa}n_3^{\lambda}) \label{kaplambulk}
\eea
As with other orbifolds, it is evident that in order to get {\em odd} intersection numbers, as required by eq. (\ref{ab3}), we shall need to make use of exceptional 3-cycles, constructed using the collapsed 2-cycles that arise in the $\theta^6$-twisted sector. 

In the $\theta ^6$-twisted sector there are 16 fixed tori $T^2_3$ at the $\mathbb{Z}_2$ fixed points $f_{\sigma_1, \sigma_2,\sigma_3,\sigma_4}$ on the $SO(8)$ lattice, where
\beq
f_{\sigma_1, \sigma_2,\sigma_3,\sigma_4}:= \frac{1}{2}\sum _{a=1}^4 \sigma _a \alpha _a
\eeq
with $\sigma _a=0,1$. 
For ease of reference, we use the same notation as in the $\mathbb{Z}_6'$ case \cite{Bailin:2006zf}, denoting the fixed points by $f_{i,j}$ with the pairs $(\sigma _1,\sigma_2)$ and $(\sigma _3,\sigma _4)$ given the labels $i,j=1,4,5,6$ respectively for the values $(0,0),(1,0),(0,1),(1,1)$.
Under the  action of the point-group the 16 fixed points split into four sets, each set  transforming into itself as follows:
\bea
&&f_{1,1}\ \ {\rm invariant} \\
&& f_{4,4} \rightarrow f_{1,6} \rightarrow f_{4,5} \rightarrow f_{4,4}  \label{f3}\\
&& f_{4,1} \rightarrow f_{6,4} \rightarrow f_{6,6} \rightarrow f_{4,6} \rightarrow f_{5,6} \rightarrow f_{5,5}\rightarrow f_{4,1} \label{f61}\\
&& f_{5,1} \rightarrow f_{6,1} \rightarrow f_{1,4} \rightarrow f_{6,5} \rightarrow f_{5,4} \rightarrow f_{1,5}\rightarrow f_{5,1} \label{f62}
\eea
There are then four non-zero invariant exceptional 3-cycles:
\bea
\epsilon _1:=(1+\theta +\theta ^2+ ... +\theta ^{5})f_{4,1} \otimes \pi _5&=& (f_{4,1}-f_{6,4}+f_{6,6}-f_{4,6}+f_{5,6}-f_{5,5}) \otimes \pi _5  \label{eps1}\\
\tilde{\epsilon} _1:=(1+\theta +\theta ^2+ ... +\theta ^{5})f_{4,1} \otimes \pi _6&=& (f_{4,1}-f_{6,4}+f_{6,6}-f_{4,6}+f_{5,6}-f_{5,5}) \otimes \pi _6  \\
 \epsilon _2 :=(1+\theta +\theta ^2+ ... +\theta ^{5})f_{5,1} \otimes \pi _5&=& (f_{5,1}-f_{6,1}+f_{1,4}-f_{6,5}+f_{5,4}-f_{1,5}) \otimes \pi _5 \\
 \tilde{\epsilon} _2 :=(1+\theta +\theta ^2+ ... +\theta ^{5})f_{5,1} \otimes \pi _6&=& (f_{5,1}-f_{6,1}+f_{1,4}-f_{6,5}+f_{5,4}-f_{1,5})\otimes \pi _6 \label{epst2}
\eea
which is consistent with the cohomology of the $\theta^6$-twisted sector. The self-intersection number of a ($\mathbb{Z}_2$) collapsed 2-cycle is, as before, given by
\beq
f_{i,j} \circ f_{k,\ell} =-2\delta_{i,k}\delta_{j,\ell}  \label{fijkl}
\eeq
Then, 
\beq
\epsilon_i \circ \tilde{\epsilon}_j=2\delta _{ij} = -\tilde{\epsilon}_i \circ \epsilon _j \quad i,j=1,2
\eeq
(The corresponding results for the $\mathbb{Z}_{12}$-I case are given in the Appendix.)
The general exceptional   brane $\Pi ^{\rm ex}_{\kappa}$ is then given by
\beq
\Pi ^{\rm ex}_{\kappa}=\sum_{i=1}^2 e^{\kappa}_i(n^{\kappa}_3 \epsilon_i+m^{\kappa}_3 \tilde{\epsilon}_i) \label{pikapex}
\eeq
where the coefficients $e_i^{\kappa}$  are determined by the fixed points wrapped by the 2-cycle used to construct $\Pi ^{\rm bulk}_{\kappa}$, as we shall see in the following section. For two general exceptional branes of this form
\beq
\Pi ^{\rm ex}_{\kappa} \circ \Pi ^{\rm ex}_{\lambda}=2(e^{\kappa}_1e^{\lambda}_1+e^{\kappa}_2e^{\lambda}_2)(n_3^{\kappa}m_3^{\lambda}-m_3^{\kappa}n_3^{\lambda}) \label{kaplamex}
\eeq

Exceptional cycles also arise in other twisted sectors. For example, in the $\theta^4$-sector there are 9 fixed tori at the $\mathbb{Z}_3$ fixed points
\beq
g_{m,p}:=\frac{1}{3}[m(\alpha _4 -\alpha _1-\alpha_3)+p(\alpha _2-\alpha_3)]
\eeq
with $m,p=0,1,2$, and, as above, collapsed 2-cycles at these fixed points may be combined with 1-cycles in $T^2_3$ to construct further twisted 3-cycles. 
However, only bulk cycles and exceptional cycles  at $\mathbb{Z}_2$ fixed points have a known interpretation in terms of partition functions \cite{Blumenhagen:2002wn} . 
In what follows we have therefore  only considered the exceptional 3-cycles defined in eqns (\ref{eps1}) ... (\ref{epst2}). 
\section{Factorisable 2-cycles} \label{Sfact}
The general 2-cycle on the $SO(8)$ lattice that appears in eq. (\ref{kappa}) has the form
\beq
\Pi_2= \sum_{a<b} n_{a,b}\pi _{a,b} \label{Pi2}
\eeq
with $a,b=1,2,...,4$ and $n_{a,b}$ six arbitrary integers. 
Now suppose that $\Pi _2$ is the product of two 1-cycles $\sum _an_a \pi_a$ and $\sum _bm_b\pi _b$, where $n_a$ and $m_b$ are integers. In this case the six integers $n_{a,b}$ are expressible in terms of the eight integers  $n_a$ and $m_b$ as 
\beq
n_{a,b}=n_am_b-m_an_b
\eeq
They then satisfy the constraint
\beq
n_{1,2}n_{3,4}+n_{1,4}n_{2,3}=n_{1,3}n_{2,4} \label{fact}
\eeq 
A general set of six wrapping numbers $n_{a,b}$ will generally {\em not} satisfy this constraint, and even if they do it is not sufficient to ensure that $\Pi_2$ is ``factorisable'' in this way. If it is, it is straightforward to identify the four fixed points $f_{i,j}$ that are wrapped by $\Pi_2$. For example, if such a factorisable 2-cycle has $(n_{1,2},n_{1,3},n_{1,4},n_{2,3},n_{2,4},n_{3,4})=(1,0,0,0,0,0) \bmod 2$, then $(n_3,n_4)=(0,0)\bmod2=(m_3,m_4)$ and either $(n_1,n_2)=(1,0) \bmod2$ and $(m_1,m_2)=(0,1) \ {\rm or} \ (1,1) \bmod2$, or {\it vice versa}. Evidently $\Pi_2$, like $\pi _{1,2}$, wraps the four fixed points $f_{1,j},f_{4,j},f_{5,j},f_{6,j}$ with $j=1,4,5,6$ arbitrary. Henceforth we shall only consider such factorisable 2-cycles.

{\it A priori}, there are $2^6$ cases to consider for the set  $(n_{1,2},n_{1,3},n_{1,4},n_{2,3},n_{2,4},n_{3,4}) \bmod 2$. However, the case in which all $n_{i,j}$ are even is of no physical interest, since we require  the wrapping numbers to have no common factor.  The action of $\theta$ splits the remaining 63 cases into sets as follows:
\beq
63=3(1)+6(2)+4(3)+6(6)
\eeq
and we only need to keep one representative of each of the 19 sets. In fact, only 9 of these can satisfy the factorisation constraint given in eq. (\ref{fact}). They are listed in Table \ref{class} together with the associated values of $a_{1,2}\bmod 2$; these are defined in eqs (\ref{a1}) and (\ref{a2}). 
\begin{table}
 \begin{center}
\begin{tabular}{||c|c||} \hline \hline
$(n_{1,2},n_{1,3},n_{1,4},n_{2,3},n_{2,4},n_{3,4})\bmod 2$ & $(a_1,a_2) \bmod2$ \\ \hline  \hline
$(0,1,1,0,0,1)$ & $(0,0)$ \\ 
\hline 
$(0,0,0,1,0,0)$ & $(1,0)$  \\ 
$(1,1,1,0,0,0)$ & $(0,1)$  \\ \hline 
$(0,0,0,0,0,1)$ & $(0,0)$ \\ 
$(0,1,1,0,0,0)$ & $(0,0)$ \\  \hline 
$(1,0,0,0, 0,0)$ & $(0,1)$ \\ 
$(0,1,0,0, 0,0)$ & $(1,1)$  \\ 
$(0,0,1,0, 0,0)$ & $(1,1)$ \\ 
$(1,1,0,0, 0,0)$ & $(1,0)$ \\ \hline \hline
\end{tabular}
\end{center} 
\caption{ \label{class}  Representatives of the 9 potentially factorisable classes of 2-cycles.}
 \end{table}

Each of these classes is associated with four sets of four fixed points, as illustrated above. The bulk part $\Pi_{\kappa}^{\rm bulk}$ of a fractional brane $\kappa$, where 
\beq
\kappa=\frac{1}{2}\Pi_{\kappa}^{\rm bulk}+\frac{1}{2}\Pi_{\kappa}^{\rm ex}, \label{kapfrac}
\eeq
 is determined by the 3-cycle given in eq. (\ref{kappa}). Supersymmetry requires that it wraps the four fixed points that determimine the exceptional part $\Pi_{\kappa}^{\rm ex}$ as follows.  
The four fixed points contribute with a sign determined by the  Wilson lines $t^{\kappa}_0, t^{\kappa}_1,t^{\kappa}_2= \pm 1$. In the example given above, the four fixed points $f_{1,1},f_{4,1},f_{5,1},f_{6,1}$ are associated with the invariant exceptional 3-cycle generated by
$t^{\kappa}_0(f_{1,1}+t^{\kappa}_2f_{4,1}+t^{\kappa}_1f_{5,1}+t^{\kappa}_1t^{\kappa}_2f_{6,1})\otimes(n^{\kappa}_3\pi _5+m^{\kappa}_3\pi_6)$, which gives 
\beq
\Pi_{\kappa}^{\rm ex}=\sum^2_{i=1}(\alpha^{\kappa}_i \epsilon _i +\tilde{\alpha}^{\kappa}_i \tilde{\epsilon} _i )
\eeq
where
\bea
\alpha^{\kappa}_i=n_3^{\kappa}e^{\kappa}_i\\
\tilde{\alpha}^{\kappa}_i=m_3^{\kappa}e^{\kappa}_i
\eea
and in this example
\bea
e^{\kappa}_1 &=&t^{\kappa}_0t^{\kappa}_2 \\
e^{\kappa}_2&=&t^{\kappa}_0t^{\kappa}_1(1-t^{\kappa}_2)
\eea
The fixed points  for all 9 classes, together with the corresponding values for $e^{\kappa}_1$ and  $e^{\kappa}_2$, are listed in Table \ref{ex}. 
\begin{table}
 \begin{center}
\begin{tabular}{||c c|c|c|c|c||} \hline \hline
$n^{\kappa}_{a,b}\bmod 2$ &$f_{i,j}$&$a^{\kappa}_1\bmod2$& $a^{\kappa}_2\bmod2$ &$e^{\kappa}_1$ & $e^{\kappa}_2$  \\ \hline \hline
$(1,0,0,0,0,0)$ &$f_{1,1}, f_{4,1},f_{5,1},f_{6,1}$&0&1&$t_2$&$t_1(1-t_2)$ \\ 
&$f_{1,4}, f_{4,4},f_{5,4},f_{6,4}$&&&$-t_1t_2$ & $1+t_1$ \\ 
I&$f_{1,5}, f_{4,5},f_{5,5},f_{6,5}$&&&$-t_1$ & $-(1+t_1t_2)$ \\ 
&$f_{1,6}, f_{4,6},f_{5,6},f_{6,6}$&&&$t_1t_2+t_1-t_2$& $0$ \\ \hline 
$(0,1,0,0,0,0)$ &$f_{1,1}, f_{4,1},f_{1,4},f_{4,4}$&1&1&$t_2$&$t_1$ \\ 
&$f_{5,1}, f_{6,1},f_{5,4},f_{6,4}$&&&$-t_1t_2$ & $1+t_1-t_2$ \\ 
II&$f_{1,5}, f_{4,5},f_{1,6},f_{4,6}$&&&$-t_1t_2$& $-1$ \\ 
&$f_{5,5}, f_{6,5},f_{5,6},f_{6,6}$&&&$t_1t_2+t_1-1$& $-t_2$ \\ \hline 
$(0,0,1,0,0,0)$ &$f_{1,1}, f_{4,1},f_{1,5},f_{4,5}$&1&1&$t_2$&$-t_1$ \\ 
&$f_{5,1}, f_{6,1},f_{5,5},f_{6,5}$&&&$-t_1$ & $1-t_2-t_1t_2$ \\ 
III&$f_{1,4}, f_{4,4},f_{1,6},f_{4,6}$&&&$-t_1t_2$& $1$ \\ 
&$f_{5,4}, f_{6,4},f_{5,6},f_{6,6}$&&&$t_1t_2+t_1-t_2$& $1$ \\ \hline 
$(0,0,0,1,0,0)$ &$f_{1,1}, f_{5,1},f_{1,4},f_{5,4}$&1&0&0&$t_1+t_2+t_1t_2$ \\ 
&$f_{4,1}, f_{6,1},f_{4,4},f_{6,4}$&&&$1-t_1t_2$ & $-t_2$ \\ 
IV&$f_{1,5}, f_{5,5},f_{1,6},f_{5,6}$&&&$t_2(t_1-1)$& $-1$ \\ 
&$f_{4,5}, f_{6,5},f_{4,6},f_{6,6}$&&&$t_1(t_2-1)$& $-t_2$ \\ \hline
$(0,0,0,0,0,1)$ &$f_{1,1}, f_{1,4},f_{1,5},f_{1,6}$&0&0&$0$&$t_2-t_1$ \\ 
&$f_{4,1}, f_{4,4},f_{4,5},f_{4,6}$&&&$1-t_1t_2$&$0$ \\ 
V&$f_{5,1}, f_{5,4},f_{5,5},f_{5,6}$&&&$t_1(t_2-1)$&$1+t_2$ \\ 
&$f_{6,1}, f_{6,4},f_{6,5},f_{6,6}$&&&$t_2(t_1-1)$&$-(1+t_1)$ \\ \hline \hline
$(1,1,0,0,0,0)$ &$f_{1,1}, f_{4,1},f_{5,4},f_{6,4}$&1&0&$t_2(1-t_1)$&$t_1$ \\ 
&$f_{1,5}, f_{4,5},f_{5,6},f_{6,6}$&&&$t_1(1+t_2)$ & $-1$ \\ 
VI&$f_{5,1}, f_{6,1},f_{1,4},f_{4,4}$&&&$0$& $1+t_1-t_2$ \\ 
&$f_{5,5}, f_{6,5},f_{1,6},f_{4,6}$&&&$-(1+t_1t_2)$& $-t_2$ \\ \hline
$(0,1,1,0,0,0)$ &$f_{1,1}, f_{4,1},f_{4,6},f_{5,4}$&0&0&$t_2(1-t_1)$&$0$ \\ 
&$f_{5,1}, f_{6,1},f_{5,6},f_{6,6}$&&&$t_1(1+t_2)$ & $1-t_2$ \\ 
VII&$f_{1,4}, f_{4,4},f_{1,5},f_{4,5}$&&&$0$& $1-t_1$ \\ 
&$f_{5,4}, f_{6,4},f_{5,5},f_{6,5}$&&&$-(t_1+t_2)$& $1-t_1t_2$ \\ \hline \hline
$(1,1,1,0,0,0)$ &$f_{1,1}, f_{4,1},f_{5,6},f_{6,6}$&0&1&$t_1+t_2+t_1t_2$&$0$ \\ 
&$f_{5,1}, f_{6,1},f_{1,6},f_{4,6}$&&&$-t_1t_2$ & $1-t_2$ \\ 
VIII&$f_{1,4}, f_{4,4},f_{5,5},f_{6,5}$&&&$-t_1$& $1-t_1t_2$ \\ 
&$f_{5,4}, f_{6,4},f_{1,5},f_{4,5}$&&&$-t_1t_2$& $t_1-1$ \\ \hline
$(0,1,1,0,0,1)$ &$f_{1,1}, f_{1,6},f_{4,5},f_{4,4}$&0&0&$0$&$0$ \\ 
&$f_{5,1}, f_{5,6},f_{6,5},f_{6,4}$&&&$t_2(1-t_1)$ & $1-t_1$ \\ 
IX&$f_{4,1}, f_{4,6},f_{1,5},f_{1,4}$&&&$1-t_2$& $t_1(1-t_2)$ \\ 
&$f_{6,1}, f_{6,6},f_{5,5},f_{5,4}$&&&$t_2-t_1$& $t_1t_2-1$ \\ \hline \hline
\end{tabular}
\end{center} 
\caption{ \label{ex}  The fixed points and coefficients $e^{\kappa}_i$ of the exceptional cycles associated with the 9 classes of factorisable 2-cycles; an overall factor of $t_0$ is omitted. }
 \end{table}

\section{Supersymmetric bulk 3-cycles}
The action  of the point group generator given in eq. (\ref{z122}) ensures that the closed-string sector is supersymmetric, but to avoid supersymmetry breaking in the open-string sector the D6-branes must wrap special Lagrange cycles. That is to say, we require that
\bea
X^{\kappa} := {\rm Re} \ \Omega|_{\Pi^{\kappa}}>0  \label{Xkap}\\
Y^{\kappa} := {\rm Im} \ \Omega|_{\Pi^{\kappa}}=0 \label{Ykap}
\eea
where
\beq
\Omega :=dz_1 \wedge dz_2 \wedge dz_3 
\eeq
is the holomorphic 3-form. The complex coordinates $z_1$ and $z_2$ are those which diagonalise the action of $\theta$ as in eq. (\ref{thetz}) with $v_1$ and $v_2$ as given in eq. (\ref{z122}). The 2-cycle $\pi _{a,b}$ may be parametrised as
\beq
\pi_{a,b}=\lambda \pi_a+\mu \pi_b \quad {\rm with}   \quad  0\leq\lambda,\mu<1
\eeq
so to evaluate $dz_1 \wedge dz_2 $ on $\pi_{a,b}$ we need a representation of the four simple roots $\alpha_a$ in this complex basis:
\beq
\alpha _a= (w_1^{(a)},w_2^{(a)})
\eeq
Defining the central root by the general form
\beq
\alpha_2=\sqrt{2}(e^{i\phi_1} \cos \theta,e^{i\phi_2} \sin \theta) \quad {\rm with}   \quad 0\leq \theta \leq \pi/2 \quad {\rm and} \quad  0\leq \phi_{1,2}<2\pi \label{alf2}
\eeq
so that $\alpha_2.\alpha_2=2$, it is easy to verify that the remaining roots are given by
\bea
\alpha _1&=&-\sqrt{2}(e^{i\phi_1}\cos \theta  (1+\beta), e^{i\phi_2}\sin \theta  (1-\beta^{-1})) \label{alf1} \\
\alpha _3&=&\sqrt{2}(-e^{i\phi_1}\cos \theta \ \beta^2 , e^{i\phi_2}\sin \theta \ \beta^4 ) \label{alf3} \\
\alpha _4&=&\sqrt{2}(e^{i\phi_1}\cos \theta \ \beta^{-1}, -e^{i\phi_2}\sin \theta \ \beta)  \label{alf4}
\eea
where $\beta :=e^{i\pi/6}$ and $\cos 2\theta=-1/\sqrt{3}$. We parametrise the 1-cycle in $T^2_3$ by
\beq
z_3=\nu(n^{\kappa}_3e_5+m^{\kappa}_3 e_6) \quad {\rm with}   \quad  0 \leq \nu <1 \label{z32}
\eeq
where $e_5$ and $e_6$ define the $SU(2)\times SU(2)$ lattice.  Then, with $\pi_{\kappa}$ as defined in eq. (\ref{kappa}), we find
\bea
\Omega|_{\pi_{\kappa}}&=&\sum_{a,b}n_{a,b}^{\kappa}(w_1^{(a)}w_2^{(b)}-w_1^{(b)}w_2^{(a)})(n^{\kappa}_3+m^{\kappa}_3 \tau_3)e_5 \ d\lambda \wedge d\mu \wedge d\nu \\
&=&\sqrt{2}e^{i(\phi_1+\phi_2)} \ e_5[iA^{\kappa}_1-A^{\kappa}_2+\tau_3(iA^{\kappa}_3-A^{\kappa}_4)]\ d\lambda \wedge d\mu \wedge d\nu \label{Ompikap}
\eea
where $\tau_3:=e_6/e_5$ is the complex structure of $T^2_3$. The phases of $e_5$ and $e_6$ as well as $\phi_1$ and $\phi_2$ are constrained by the requirement that the orientifold embedding  of the world-sheet parity operator also acts as an automorphism of the lattice.
\section{The $\mathbb{Z}_{12}$-II orientifold} \label{RII}
The embedding $\mathcal{R}$ of the world-sheet  parity operator acts on the three complex coordinates $z_k$ as complex conjugation
\beq
\mathcal{R}z_k=\bar{z}_k \quad (k=1,2,3) \label{Rzk}
\eeq
In particular, since we require that \R \ acts crystallographically on the root lattice, this requires that
\beq
\mathcal{R} \alpha _a= \bar{\alpha}_a=\sum _bN_{ab} \alpha _b
\eeq
where  $N_{ab} \in \mathbb{Z}$. This leads to six independent solutions which are displayed in Table \ref{R2}.
\begin{table}
 \begin{center}
  \resizebox{\textwidth}{!}{%
\begin{tabular}{||c||c|c|c|c||c|c||} \hline \hline
{\bf Lattice}&\R$\alpha_1$&\R$\alpha _2$&\R$\alpha _3$&\R$\alpha _4$& $e^{-2i\phi_1}$& $e^{-2i\phi_2}$ \\ \hline \hline
{\bf a}&$-(\alpha _2+\alpha _4)$&$ \alpha _2$&$-(\alpha_2+\alpha_3)$& $-(\alpha_1+\alpha_2)$& $1$ & $1$ \\
{\bf b}&$-(\alpha _1+\alpha _2+\alpha _3+\alpha _4)$&$\alpha _1+ \alpha _2+ \alpha _4$& $ -(\alpha_1+\alpha_2)$& $\alpha_2+\alpha_3$& $- \beta^3$ & $- \beta^3$ \\
{\bf c}&$-\alpha_1$&$\alpha _1+ \alpha _2$&$\alpha_4$  & $\alpha_3$& $- \beta$ & $ \beta^{-1}$ \\
{\bf d}&$-(\alpha _2+\alpha _3+\alpha _4)$&$ \alpha _4$&  $-(\alpha_1+\alpha_2+\alpha_4)$    &    $\alpha_2$& $ \beta^{-1}$ & $- \beta$ \\
{\bf e}&$-(\alpha _1+\alpha _2+\alpha _3)$&$\alpha _3$ &$\alpha_2$   &$\alpha_1+\alpha_2+\alpha_4$& $- \beta^{2}$ & $- \beta^{-2}$ \\
{\bf f}&$-(\alpha _1+2\alpha _2+\alpha _3+\alpha _4)$&$\alpha _2+ \alpha _3$& $-\alpha_3$&            $\alpha_4$&$ \beta^{-2}$ & $ \beta^2$ \\ 
\hline \hline
\end{tabular}}
\end{center} 
\caption{ \label{R2}  The action of \R \ and the phases $\phi _1$ and $\phi _2$ for crystallographic action of \R \ on $\alpha_a \ (a=1,2,3,4)$; an overall sign of $\epsilon= \pm 1$ is undisplayed.}
 \end{table}
For the bulk 3-cycles $\rho_p \ (p=1,2,...,4)$ defined in eqs (\ref{ro1})-(\ref{ro4}), only two combinations $\sigma _{1,2}$  of 2-cycles enter the invariant bulk 3-cycles:
\bea
\sigma_1&:=&\pi _{1,4}+\pi _{2,3}+\pi _{2,4} \\
\sigma_2&:=&-\pi _{1,3}-\pi _{2,3}+\pi _{2,4}+\pi _{3,4}
\eea
It is easy to verify that the six different lattices reduce to just two classes when acting on these combinations:
\bea
({\bf a,e,f}): \quad \mathcal{R} \sigma _1=-\sigma_1, \quad  \mathcal{R} \sigma _2=\sigma_2\\
({\bf b,c,d}): \quad \mathcal{R} \sigma _1=\sigma_1, \quad  \mathcal{R} \sigma _2=-\sigma_2
\eea
Note too that, independently of the overall sign $\epsilon$, the product of the phases given in Table \ref{R2} restricts the hitherto unknown phase in eq. (\ref{Ompikap}) 
\bea
( {\bf a,e,f}): \quad e^{i(\phi_1+\phi _2)}= \pm1 \label{aef} \\
(  {\bf b,c,d}):  \quad e^{i(\phi_1+\phi _2)}= \pm i \label{bcd}
\eea
As in the $\mathbb{Z}_6'$ case, the action of \R \ on the basis 1-cycles $\pi _{5,6}$ in $T^2_3$ is given by
\bea
{\bf A}: \quad \mathcal{R} \pi _5=\pi _5, \quad \  \mathcal{R}\pi _6=-\pi _6 \label{A}\\
 {\bf B}: \quad \mathcal{R} \pi _5=\pi _5, \quad \  \mathcal{R}\pi _6=\pi _5-\pi _6 \label{B}
\eea
Thus, in both cases $e_5$ is real and chosen to be positive, and the complex structure of $T^2_3$ is given by
\beq
\tau_3=b+i{\rm Im} \ \tau_3
\eeq
with $b=0$ or $b=1/2$ respectively for the {\bf A} and {\bf B} lattices.
Hence there are just four different classes of behaviour of the bulk 3-cycles under the action of \R. 
The results are displayed in Table \ref{Rrho}.
\begin{table}
 \begin{center}
\begin{tabular}{||c||c|c|c|c||} \hline \hline
{\bf Lattice} & $\mathcal{R}\rho _1$ & $\mathcal{R}\rho _2$ &$\mathcal{R}\rho _3$ &$\mathcal{R}\rho _4$ \\ \hline
{\bf (a,e,f)A} & $-\rho_1$ &  $\rho_2$ & $\rho_3$ & $-\rho_4$  \\ 
{\bf (a,e,f)B} & $-\rho_1$ &  $\rho_2$ & $-\rho_1+\rho_3$ & $\rho_2-\rho_4$  \\ \hline
{\bf (b,c,d)A} & $\rho_1$ &  $-\rho_2$ & $-\rho_3$ & $\rho_4$  \\ 
{\bf (b,c,d)B} & $\rho_1$ &  $-\rho_2$ & $\rho_1-\rho_3$ & $-\rho_2+\rho_4$
  \\  \hline \hline
\end{tabular}
\end{center} 
\caption{ \label{Rrho}  The action of \R \ on the invariant 3-cycles.}
 \end{table}
Choosing the lower signs in eqs (\ref{aef}) and (\ref{bcd}),  the functions $X^{\kappa}$ and $Y^{\kappa}$ defined in eqs (\ref{Xkap}) and (\ref{Ykap}) are then given in Table \ref{XY}.
\begin{table}
 \begin{center}
\begin{tabular}{||c||c|c||} \hline \hline
{\bf Lattice} & $X^{\kappa}$ & $Y^{\kappa}$ \\ \hline\hline
{\bf (a,e,f) A}& $A_2^{\kappa}+ {\rm Im} \ \tau _3 A^{\kappa}_3$ & $-A_1^{\kappa}+ {\rm Im} \ \tau _3 A^{\kappa}_4$ \\
{\bf (a,e,f) B}& $A_2^{\kappa}+\frac{1}{2}A_4^{\kappa}+ {\rm Im} \ \tau _3 A^{\kappa}_3$ & $-A_1^{\kappa}-\frac{1}{2}A_3 ^{\kappa}+ {\rm Im} \ \tau _3 A^{\kappa}_4$ \\ \hline
{\bf (b,c,d) A}&$A_1^{\kappa}- {\rm Im} \ \tau _3 A^{\kappa}_4$& $A_2^{\kappa}+ {\rm Im} \ \tau _3 A^{\kappa}_3$   \\
{\bf (b,c,d) B}&$A_1^{\kappa}+\frac{1}{2}A_3^{\kappa}- {\rm Im} \ \tau _3 A^{\kappa}_4$& $A_2^{\kappa}+\frac{1}{2}A_4^{\kappa}+ {\rm Im} \ \tau _3 A^{\kappa}_3$ \\
\hline \hline
\end{tabular}
\end{center} 
\caption{ \label{XY}  The functions $X^{\kappa}$ and $Y^{\kappa}$. (A global positive factor of $\sqrt{2} e_5$ for each entry is omitted).}
 \end{table}


As already noted, the orientifold action leads to the formation of O6-planes. To determine these we must first identify the two \R- and two $\theta$\R-invariant 1-cycles on each configuration of the $SO(8)$ lattice.  These are displayed in Table  \ref{Rtheta1cyc}, as is the single \R- and single $\theta$\R-invariant 1-cycle on $T^2_3$. The corresponding \R- and  $\theta$\R-invariant 3-cycles then generate the bulk 3-cycles displayed in Table \ref{o6}; the  overall sign is fixed by the supersymmetry requirement that $X^{\kappa}$ is positive. 
\begin{table}
 \begin{center}
\begin{tabular}{||c||c|c||} \hline \hline
{\bf Lattice} & Invariant & 1-cycle(s) \\ \hline \hline
$SO(8)${\bf a} & \R&$\pi_2, \  \pi _1-\pi_4$ \\
&$\theta$\R&$\pi_1, \ \pi_3-\pi_4$ \\
$SO(8)${\bf b} & \R&$\pi_1+\pi_2-\pi_3, \  \pi _2+\pi_3+\pi_4$ \\
&$\theta$\R&$\pi_4, \ 2\pi_2+\pi_3$ \\
$SO(8)${\bf c} & \R&$\pi_1+2\pi_2, \  \pi _3+\pi_4$ \\
&$\theta$\R&$\pi_1-\pi_3+2\pi_4, \ \pi_2+\pi_3$ \\
$SO(8)${\bf d} & \R&$\pi_1-\pi_3, \  \pi _2+\pi_4$ \\
&$\theta$\R&$\pi_2, \ \pi_1-\pi_4$ \\
$SO(8)${\bf e} & \R&$\pi_1-\pi_3+2\pi_4, \  \pi _2+\pi_3$ \\
&$\theta$\R&$\pi_1+\pi_2-\pi_3, \  \pi_2+\pi_3+\pi_4$ \\
$SO(8)${\bf f} & \R&$\pi_4, \  2\pi _2+\pi_3$ \\
&$\theta$\R&$\pi_1-\pi_3, \ \pi_2+\pi_4$ \\ \hline
$T^2_3${\bf A} & \R& $\pi_5$ \\
&$\theta$\R &$\pi_6$ \\

$T^2_3${\bf B} & \R& $\pi_5$ \\
&$\theta$\R &$\pi_5-\pi_6$ \\
\hline \hline
\end{tabular}
\end{center} 
\caption{ \label{Rtheta1cyc}  \R- and $\theta$\R-invariant 1-cycles.}
 \end{table}
\begin{table}
 \begin{center}
\begin{tabular}{||c||c|c|c||} \hline \hline
{\bf Lattice} & Invariant &$(n_{1,2},n_{1,3},n_{1,4},n_{2,3},n_{2,4},n_{3,4})(n_3,m_3)$&  3-cycle \\ \hline \hline
{\bf aA} & \R & $(1,0,0,0,1,0)(1,0)$ & $2\rho _2$ \\
             & $\theta$\R& $(0,1,-1,0,0,0)(0,1)$&$2s\rho _3$ \\
{\bf aB} & \R & $(1,0,0,0,1,0)(1,0)$ & $2\rho _2$ \\
             & $\theta$\R& $(0,1,-1,0,0,0)(1,-1)$&$2s(-\rho_1+2\rho _3)$ \\ \hline
{\bf bA} & \R & $(1,1,1,1,1,-1)(1,0)$ & $2\rho _1$ \\
             & $\theta$\R& $(0,0,0,0,2,1)(0,1)$&$-2s\rho _4$ \\
{\bf bB} &  \R & $(1,1,1,1,1,-1)(1,0)$ & $2\rho _1$ \\
             & $\theta$\R& $(0,0,0,0,2,1)(1,-1)$&$2s(\rho_2-2\rho _4)$ \\ \hline
 {\bf cA} & \R & $(0,1,1,2,2,0)(1,0)$ & $2\rho _1$ \\
             & $\theta$\R& $(1,1,0,1,-2,-2)(0,1)$&$-2s\rho _4$ \\
 {\bf cB} & \R & $(0,1,1,2,2,0)(1,0)$ & $2\rho _1$ \\
             & $\theta$\R& $(1,1,0,1,-2,-2)(1,-1)$&$2s(\rho_2-2\rho _4)$ \\ \hline
{\bf dA} & \R & $(1,0,1,1,0,-1)(1,0)$ & $2\rho _1$ \\
             & $\theta$\R& $(1,0,0,0,1,0)(0,1)$&$-2s\rho _4$ \\
{\bf dB} & \R & $(1,0,1,1,0,-1)(1,0)$ & $2\rho _1$ \\
             & $\theta$\R& $(1,0,0,0,1,0)(1,-1)$&$2s(\rho_2-2\rho _4)$ \\ \hline
{\bf eA} & \R & $(1,1,0,1,-2,-2)(1,0)$ & $2\rho _2$ \\
             & $\theta$\R& $(1,1,1,2,1,-1)(0,1)$&$2s\rho _3$ \\
{\bf eB} & \R & $(1,1,0,1,-2,-2)(1,0)$ & $2\rho _2$ \\
             & $\theta$\R& $(1,1,1,2,1,-1)(1,-1)$&$2s(-\rho_1+2\rho _3)$ \\ \hline
{\bf fA} & \R & $(0,0,0,0,2,1)(1,0)$ & $2\rho _2$ \\
             & $\theta$\R& $(1,0,1,1,0,-1)(0,1)$&$2s\rho _3$ \\
{\bf fB} & \R & $(0,0,0,0,2,1)(1,0)$ & $2\rho _2$ \\
             & $\theta$\R& $(1,0,1,1,0,-1)(1,-1)$&$2s(-\rho_1+2\rho _3)$ \\
 \hline \hline
\end{tabular}
\end{center} 
\caption{ \label{o6}  Supersymmetric \R- and $\theta$\R-invariant bulk 3-cycles of the $\mathbb{Z}_{12}$-II  orientifold; $s=\pm 1$ is the sign of ${\rm Im} \ \tau_3$.}
 \end{table} 
The O6-plane is then the sum of the two orbits, which gives:
\bea
{\bf (a,e,f)A}: && \pi_{\rm O6}=2(\rho_2+s\rho_3) \\
{\bf (a,e,f)B}: && \pi_{\rm O6}=2[\rho_2+s(-\rho_1+2\rho_3)] \label{o6aefB}\\
{\bf (b,c,d)A}: && \pi_{\rm O6}=2(\rho_1-s\rho_4) \\
{\bf (b,c,d)B}: && \pi_{\rm O6}=2[\rho_1+s(\rho_2-2\rho_4)] \label{o6bcdB}
\eea
where $s$ is the sign of ${\rm Im} \ \tau_3$.

We also need the action of \R \ on the exceptional cycles $\epsilon _j$ and $\tilde{\epsilon}_j$, which in turn depends upon the action of \R \ on the sixteen  $\mathbb{Z}_2$ fixed points $f_{i,j} \ (i,j=1,4,5,6)$ in the $\theta^6$-twisted sector. This may be determined using the action of \R \ on the simple roots $\alpha_a$ of the $SO(8)$ lattice, which is displayed in Table  \ref{R2}. On all six lattices there are 4 invariant fixed points and 6 pairs that transform into each other under the action of \R. These are displayed in Table \ref{Rfij}.
\begin{table}
 \begin{center}
\begin{tabular}{||c||c|c||} \hline \hline
{\bf Lattice} & Invariants & Pairs \\ \hline \hline
{\bf a} & $f_{1,1},f_{5,1}, f_{4,5}, f_{6,5}$ & $ (f_{4,1},f_{5,5}),(f_{6,1},f_{1,5}),(f_{1,4},f_{5,4}) ,     (f_{1,6},f_{4,4}), (f_{6,4},f_{5,6}),(f_{6,6},f_{4,6})$ \\
{\bf b} & $f_{1,1},f_{5,6}, f_{4,5}, f_{6,4}$ & $ (f_{4,1},f_{6,6}),(f_{5,1},f_{6,5}),(f_{6,1},f_{1,4}) ,     (f_{1,6},f_{4,4}), (f_{4,6},f_{5,5}),(f_{1,5},f_{5,4})$ \\
{\bf c} & $f_{1,1},f_{4,1}, f_{1,6}, f_{4,6}$ & $ (f_{1,4},f_{1,5}),(f_{4,4},f_{4,5}),(f_{5,4},f_{6,5}) ,     (f_{5,5},f_{6,4}), (f_{5,6},f_{6,6}),(f_{5,1},f_{6,1})$ \\
{\bf d} & $f_{1,1},f_{4,4}, f_{5,5}, f_{6,6}$ & $ (f_{1,4},f_{6,5}),(f_{1,5},f_{5,1}),(f_{1,6},f_{4,5}) ,     (f_{4,1},f_{5,6}), (f_{6,1},f_{5,4}),(f_{4,6},f_{6,4})$ \\
{\bf e} & $f_{1,1},f_{4,4}, f_{5,4}, f_{6,1}$ & $ (f_{1,4},f_{5,1}),(f_{1,5},f_{6,5}),(f_{1,6},f_{4,5}) ,     (f_{4,1},f_{6,4}), (f_{5,6},f_{4,6}),(f_{5,5},f_{6,6})$ \\
{\bf f} & $f_{1,1},f_{1,4}, f_{1,5}, f_{1,6}$ & $ (f_{4,1},f_{4,6}),(f_{5,1},f_{5,4}),(f_{6,1},f_{6,5}) ,     (f_{4,5},f_{4,4}), (f_{5,5},f_{5,6}),(f_{6,6},f_{6,4})$ \\
\hline \hline
\end{tabular}
\end{center} 
\caption{ \label{Rfij}  Action of \R \ on the $\theta ^6$-sector fixed points $f_{i,j} \ (i,j=1,4,5,6)$.}
 \end{table}
The action of \R \ on the exceptional cycles then follows from their definition in eqs (\ref{eps1}) ... (\ref{epst2}) using eqs (\ref{A}) and (\ref{B}). It is important to include also the further minus sign as detailed in eqn (4.3) of Blumenhagen {\it et al.} \cite{Blumenhagen:2002wn}; this is most easily seen by considering the action of $\mathcal{R}$ on the K\"ahler form $J:=idz_k \wedge d\bar{z}_k$. The results are displayed in Table \ref{Reps}.
\begin{table}
 \begin{center}
\begin{tabular}{||c||c|c|c|c||} \hline \hline
{\bf Lattice} & \R$\epsilon_1$ & \R$\epsilon_2$& \R$\tilde{\epsilon}_1$&\R$\tilde{\epsilon}_2$ \\ \hline
{\bf (a,e,f)A} & $\epsilon_1$&$-\epsilon_2$&$-\tilde{\epsilon}_1$&$\tilde{\epsilon}_2$ \\
{\bf (a,e,f)B} & $\epsilon_1$&$-\epsilon_2$&$\epsilon_1-\tilde{\epsilon}_1$&$-\epsilon_2+\tilde{\epsilon}_2$ \\ \hline
{\bf (b,c,d)A} & $-\epsilon_1$&$\epsilon_2$&$\tilde{\epsilon}_1$&$-\tilde{\epsilon}_2$ \\
{\bf (b,c,d)B} & $-\epsilon_1$&$\epsilon_2$&$-\epsilon_1+\tilde{\epsilon}_1$&$\epsilon_2-\tilde{\epsilon}_2$ \\ 
\hline \hline
\end{tabular}
\end{center} 
\caption{ \label{Reps}  Action of \R \ on the invariant exceptional 3-cycles $\epsilon_j$ and $\tilde{\epsilon}_j$.}
 \end{table}
\section{Fractional branes} \label{fracbranes}
As noted earlier, in order to obtain stacks which intersect at an {\em odd} number of points it is necessary to use fractional branes  of the form
given in eq. (\ref{kapfrac}), 
where  the bulk part $ \Pi ^{\rm bulk}_{\kappa}$ is of the form given in eq. (\ref{pikapbulk}), and determined by the 2-cycle wrapping numbers $n_{a,b}^{\kappa}$ and the 1-cycle wrapping numbers $(n_3^{\kappa},n_3^{\kappa})$ on $T^2_3$. The exceptional part $ \Pi ^{\rm ex}_{\kappa}$ is of the form given in eq. (\ref{pikapex}), in which, to ensure supersymmetry, the coefficients $e^{\kappa}_i$ are determined in the manner described in \S \ref{Sfact} by the fixed points $f^{\kappa}_{i,j}$ on the $SO(8)$ lattice that are wrapped by the bulk 2-cycle. 
It follows from eqs (\ref{kaplambulk}) and (\ref{kaplamex}) that
\beq
a \circ b= \left[\frac{3}{2}(a_1^{a}a_1^{b}+a_2^{a}a_2^{b})+ \frac{1}{2}(e^{a}_1e^{b}_1+e^{a}_2e^{b}_2)\right](n_3^{a}m_3^{b}-m_3^{a}n_3^{b})
\eeq
Similarly, using the results given in Tables \ref{Rrho} and \ref{Reps}, on the {\bf (a,e,f)A} lattice we find that
\beq
a \circ b'= \left[\frac{3}{2}(a_1^{a}a_1^{b}-a_2^{a}a_2^{b})+ \frac{1}{2}(-e^{a}_1e^{b}_1+e^{a}_2e^{b}_2)\right](n_3^{a}m_3^{b}+m_3^{a}n_3^{b})
\eeq
Hence
\beq
a \circ b-a \circ b'=n_3^{a}m_3^{b}(3a_2^{a}a_2^{b}+e^{a}_1e^{b}_1)-m_3^{a}n_3^{b}(3a_1^{a}a_1^{b}+e^{a}_2e^{b}_2) \label{abab1aefA}
\eeq
Now, by inspection of Table \ref{ex} we see that in all cases
\beq
e^{\kappa}_1= a^{\kappa}_2 \bmod 2 \quad {\rm and} \quad e^{\kappa}_2= a^{\kappa}_1 \bmod 2 \label{e12a12}
\eeq
Thus, on the  {\bf (a,e,f)A} lattice
\beq
 a \circ b- a \circ b'=0 \bmod 2
\eeq
Since $a\circ b+a \circ b'=(a \circ b-a \circ b') \bmod 2$, we can{\em not} satisfy eq. (\ref{ab3}). 
It is apparent from  Tables \ref{Rrho} and \ref{Reps} that on the  {\bf (b,c,d)A} lattice the orientifold image $b'$ differs only by an overall sign from that on the  {\bf (a,e,f)A} lattice. Thus the expression on the right-hand side of eq. (\ref{abab1aefA}) applies to $a \circ b+a \circ b'$ on the {\bf (b,c,d)A} lattice. Hence we  can{\em not} satisfy eq. (\ref{ab3}) on this lattice either.

Proceeding similarly, on the {\bf (a,e,f)B} lattice we find instead that
\beq
a \circ b-a \circ b'=-\frac{1}{2}m_3^{a}m_3^{b}(a_1^{a}a_1^{b}-a_2^{a}a_2^{b}+e^{a}_1e^{b}_1-e^{a}_2e^{b}_2) \bmod 2
\eeq
It follows from eq. (\ref{e12a12})  that
\beq
X_{a,b}:= a_1^{a}a_1^{b}-a_2^{a}a_2^{b}+e^{a}_1e^{b}_1-e^{a}_2e^{b}_2= 0 \bmod 2
\eeq
so to ensure that $a \circ b - a \circ b'=1 \bmod 2$, we require that
\bea
m_3^{a}= 1 \bmod 2=m_3^{b} \label{m3ab}\\
X_{a,b}= 2 \bmod 4 \label{Xab24}
\eea
For the reasons given above,  the same conclusions apply in the case of the {\bf (b,c,d)B} lattice.
The general solution of eq. (\ref{Xab24}) is given by 
\beq
(a^a_1a^b_1,a^a_2a^b_2,e^a_1e^b_1,e^a_2e^b_2)=(x,y,y,x+2)\quad {\rm or}\quad(x,y,y+2,x) \bmod 4  \label{Xab242}
\eeq
with $x,y=0,1,2,3 \bmod 4$.

Besides the requirements of supersymmetry and factorisability discussed earlier, there are two further constraints that must be imposed upon the non-abelian stacks $a$ and $b$. The first derives from the fact that on an orientifold   chiral matter in the symmetric ${\bf S}_{\kappa}$ and antisymmetric ${\bf A}_{\kappa}$ representations of the gauge group may arise at the interesections of any stack $\kappa$ with its orientifold image $\kappa '$. The dimensionality of these is given by
\bea
[{\bf S}_{\kappa}]:= ({\bf N}_{\kappa} \times {\bf N}_{\kappa})_{\rm symm}=\frac{1}{2} N_{\kappa}(N_{\kappa}+1) \\
\left[{\bf A}_{\kappa}\right]:=({\bf N}_{\kappa} \times {\bf N}_{\kappa})_{\rm antisymm}=\frac{1}{2} N_{\kappa}(N_{\kappa}-1)
\eea
Thus, on the $U(3)$ stack $a$, this gives unobserved symmetric 6-dimensional representations. Likewise, on the $U(2)$ stack $b$ unobserved 3-dimensional  chiral representations may arise. Clearly, we must demand the absence of such symmetric representations on both of  these stacks. The antisymmetric representation on the $a$ stack is the $\bar{\bf 3}$ representation. In principle such states are acceptable as quark singlets $q^c_L$ states, provided that  the hypercharge $Y(q^c_L)=2y_a$ is right. Evidently, this require that $y_a=1/6$ or $-1/3$, corresponding respectively to $d^c_L$ and $u^c_L$ states.  On the $b$ stack the antisymmetric representation is the singlet representation. Again, such states are acceptable as charged lepton singlets $\ell^c_L$, provided that $y_b=1/2$, or as neutrino singlets $\nu^c_L$, if $y_b=0$. It follows from the considerations at the end of \S 1 that only $(y_a,y_b)=(1/6,0)$ or $(-1/3,1/2)$ are consistent with getting the correct weak hypercharge for the quark doublets.
The numbers of such chiral representations are given by
\bea
\#({\bf S}_{\kappa})=\frac{1}{2}(\kappa \circ \kappa '-\kappa \circ \pi _{\rm O6}) \\
\#({\bf A}_{\kappa})=\frac{1}{2}(\kappa \circ \kappa '+\kappa \circ \pi _{\rm O6}) 
\eea
Since we must demand the absence of the symmetric ${\bf S}_a$ and ${\bf S}_b$ representations, the numbers of surviving anti-symmetric representations are
\beq
\#({\bf A}_{\kappa})=\kappa \circ \pi _{\rm O6}  \quad \kappa=a,b
\eeq
So the first additional constraint is that 
\beq
|\#({\bf A}_{\kappa})| \leq 3  \quad \kappa=a,b \label{Akap3}
\eeq
since there are only 3 quark singlets  and 3 lepton singlets of each flavour in the Standard Model.
It follows from eqs (\ref{o6aefB}) and (\ref{o6bcdB}), using the supersymmetry constraint $Y^{\kappa}=0$, with the forms of $Y^{\kappa}$ as displayed in Table \ref{XY},  that  
\bea
{\bf (a,e,f)B} \qquad \#({\bf A}_{\kappa})&=& 6[s(A^{\kappa}_3+2 A^{\kappa}_1)-A^{\kappa}_4] =6(2|{\rm Im} \ \tau_3|-1)A^{\kappa}_4 \label{AkaefB}\\
{\bf (b,c,d)B} \qquad \qquad \quad &=& -6[s(A^{\kappa}_4+2 A^{\kappa}_2)+A^{\kappa}_3]  =6(2|{\rm Im} \ \tau_3|-1)A^{\kappa}_3
\eea
Since the bulk wrapping numbers $A_p^{\kappa}$ are all integers, it is evident from the middle equations that $\#({\bf A}_{\kappa})=0 \bmod 6$. Thus,  we can{\em not} satisfy eq. (\ref{Akap3}) unless $ \#({\bf A}_{a})=0= \#({\bf A}_{b})$. On both  lattices and both stacks this requires that $A^{\kappa}_3=A^{\kappa}_4 \bmod 2$. It follows from eq. (\ref{m3ab}) that this in turn requires that 
\beq
a^{\kappa}_1=a^{\kappa}_2 \bmod 2 \quad \kappa=a,b 
\eeq
 on both lattices. If $|{\rm Im} \ \tau_3| \neq 1/2$, then on both stacks and on both lattices $(a^{\kappa}_1,a^{\kappa}_2)=(0,0) \bmod 2$, and all terms on the left-hand side of  eq. (\ref{Xab242}) are $0 \bmod 4$ so cannot satisfy eq. (\ref{Xab24}). The alternative is to require that
\beq
|{\rm Im} \ \tau_3| = \frac{1}{2}
\eeq
The solutions given in eq. (\ref{Xab242}) are now restricted to the form
\beq
(a^a_1a^b_1,a^a_2a^b_2,e^a_1e^b_1,e^a_2e^b_2)=(\underline{x,x,x,x+2}) \bmod 4 \label{Xab243}
\eeq
with $x=0,1,2,3 \bmod 4$; the underlining signifies any permutation of the underlined entries.  This can only be satisfied if at most one of $\kappa=a$ or $b$ has $(a^{\kappa}_1, a^{\kappa}_2)=(0,0) \bmod 2$. Furthermore, if, say,  $(a^{a}_1, a^{a}_2)=(0,0) \bmod 2$, and $(a^{b}_1, a^{b}_2)=(1,1) \bmod 2$, then
\beq
(a^a_1a^b_1,a^a_2a^b_2,e^a_1e^b_1,e^a_2e^b_2)=(a^a_1,a^a_2,e^a_1,e^a_2) \bmod 4 \label{Xab244}
\eeq
and eq. (\ref{Xab243}) requires that only an {\em odd} number of $a^a_1,a^a_2,e^a_1,e^a_2$ can be $2 \bmod 4$. However, in this case it is easy to verify that $a \circ a' \neq 0$, and hence  $\#({\bf S}_{a})\neq 0$. The conclusion is that only if $(a^{\kappa}_1, a^{\kappa}_2)=(1,1) \bmod 2$  for both stacks $\kappa=a$ and $b$ can this constraint be satisfied if we allow only the \SM spectrum.

Should we succeed in finding supersymmetric (factorisable)  stacks $a$ and $b$ satisfying the constraints detailed above,  it is desirable that the 
the (four-dimensional) $SU(3)$ and $SU(2)$ gauge couplings strengths unify, {\it i.e.}
\beq
\alpha _a=\alpha _b
\eeq
although we do not impose this as a constraint. 
For the gauge group $U(N_{\kappa})$, the four-dimensional fine structure constant $\alpha _{\kappa}$ of a stack $\kappa$ of $N_{\kappa}$ D6-branes wrapping a 3-cycle $\pi_{\kappa}$ is given by \cite{Klebanov:2003my,Blumenhagen:2003jy}
\beq
\frac{1}{\alpha _{\kappa}}=\frac{m_{\mathbb{P}}}{2\sqrt{2}m_{\rm string}}\frac{{\rm Vol}(\pi_{\kappa})}{\sqrt{{\rm Vol}(Y)}} 
\eeq
where $m_{\mathbb{P}}$ is the Planck mass, and $Y=T^6/\mathcal{R} \times\mathbb{Z}_{12}$-II is the compactified space in this case.
  For fractional branes $\kappa$ as defined in eq. (\ref{kapfrac})
\beq
{\rm Vol} (\kappa)= \frac{1}{2}{\rm Vol} (\Pi_{\kappa}^{\rm bulk})+ \frac{1}{2}{\rm Vol} (\Pi_{\kappa}^{\rm ex})  \simeq \frac{1}{2}{\rm Vol} (\Pi_{\kappa}^{\rm bulk})
\eeq
since the consistency of
 the supergravity approximation requires that the contribution of the bulk part
 is large compared to the contribution from the exceptional part. Then, as shown in \cite{Bailin:2011am}, for supersymmetric stacks
\bea
\frac{\alpha_a}{\alpha_b}&=& \frac{{\rm Vol}(\Pi^{\rm bulk}_{b})}{{\rm Vol}(\Pi^{\rm bulk}_{a})} \\
&=& \frac{X^b}{X^a}
\eea
where $X^{\kappa}$ is defined in eq. (\ref{Xkap}) and for the various lattices takes the values  displayed in Table \ref{XY}.
\section{Computations}
We have shown in \S \ref{fracbranes}  that the only  way that we might satisfy all of the constraints is if $a^{\kappa}_1$ and $a^{\kappa}_2$  are both odd for both stacks, {\it i.e.} if they are  of type II or III in Table \ref{ex};  then $x$ in eq. (\ref{Xab243}) is odd. The numerical search produced no solutions  satisfying the constraints in which $(a \circ b, a \circ b')=(1,2)$ or $(2,1)$. The only solutions that satisfy eq. (\ref{ab3}) (with $(a \circ b, a \circ b')=(0,3)$ or $(3,0)$) and the constraints have the wrapping numbers $(n^{\kappa}_3,m^{\kappa}_3)$ of $T^2_3$ equal to $(0, \pm 3)$ for one of the stacks, {\it i.e.}  the wrapping numbers  are not coprime; such solutions are unacceptable. The conclusion is that the \Z$_{12}$-II
 orientifold can{not} yield {\em just} the spectrum of the supersymmetric Standard Model.

 Since there are no solutions with just the supersymmetric \SM spectrum, it is of interest to  study models that approximate to it.  Instead of demanding that  $\#({\bf A}_{\kappa})=0$ for both stacks, suppose that we allow  just one, $a$ say, to have $|\#({\bf A}_{a})|=|a \circ \pi _{\rm O6}|=6$, the minimal non-zero number. On the {\bf (a,e,f)B} lattice, it then follows from eq. (\ref{AkaefB}) that
\beq
|{\rm Im} \ \tau_3|=\frac{A^a_4+\epsilon}{2A^a_4} \label{imtau3}
\eeq
where $\epsilon= \pm 1$. 
Further, since $A^a_3-A^a_4=1 \bmod 2$, it follows that $(a_1,a_2)=(1,0) \ {\rm or} \ (0,1) \bmod 2$. Thus $a$ is of type I/VIII or of type IV/VI in Table \ref{ex}. 
For the other stack, it follows that
\beq
\#({\bf A}_{b})=b \circ \pi _{\rm O6}=\frac{A^b_4}{A^a_4} \epsilon
\eeq
So if  there are no antisymmetric representations on this stack,  we require that 
\beq 
A^b_4=0=a^b_2
\eeq
Hence $A^b_2=0$ too. Also, since $2A^b_1+A^b_3=0=(2n^b_3+m^b_3)a^b_1$, it follows that $A^b_1=0=A^b_3$. This means that $X^b=0$,  which gives an infinite value for the gauge coupling strength $\alpha_b$. 
We are therefore compelled to have antisymmetric matter on {\em both} stacks. If we also require the minimal amount on $b$ too, then the stack $b$ must be of the same type as $a$ with
\beq
|A^b_4|=|A^a_4|
\eeq

Similarly, on the {\bf (b,c,d)B} lattice, if  $\#({\bf A}_{a})=6\epsilon$, then
\beq
s(2A^a_4+A^a_2)+A^a_3=(1-2|{\rm Im} \ \tau_3|)A^a_3=-\epsilon
\eeq
Hence
\beq
|{\rm Im} \ \tau_3|=\frac{A^a_3+\epsilon}{2A^a_3}
\eeq
Again, if  we demand that $\#({\bf A}_{b})=0$, then $A^b_p=0 \   (p=1,2,3,4)$, and $\alpha_b$ is infinite. Likewise,  if instead we require the minimal amount on $b$ too, then it must be of the same type as $a$ with
\beq
|A^b_3|=|A^a_3|
\eeq

Solutions for $a$ and $b$ satisfying even these weaker constraints are fairly limited. For example, on the {\bf (a,e,f)B} lattice, when both $a$ and $b$ are of type I, we find solutions of the required type with
\bea
(a^a_1,a^a_2)=(2x^a,y^a), \quad (n^a_3,m^a_3)=(0,y^a), \quad (e^a_1,e^a_2)=(z^a,2t^a) \\
(a^b_1,a^b_2)=(2x^b,y^b), \quad (n^b_3,m^b_3)=(y^b,-y^b), \quad (e^a_1,e^a_2)=(z^b,2t^b)
\eea
where $x^{\kappa}, y^{\kappa}, z^{\kappa},t^{\kappa}= \pm 1$. Then
\bea
A^a_p=(0,0,2x^ay^a,1), && (\alpha^a_i,\tilde{\alpha}^a_i)=(0,0,y^az^a,2y^at^a) \\
A^b_p=(2x^by^b,1,-2x^by^b,-1), && (\alpha^b_i,\tilde{\alpha}^b_i)=(y^bz^b,2y^bt^b,-y^bz^b,-2y^bt^b)
\eea 
and
\bea
x^ay^a=&{\rm Im} \ \tau _3&=-x^by^b \\
X^a=&\frac{5}{2}&=X^b
\eea
Then from eq. (\ref{AkaefB}), it follows that
\beq
\#({\bf A}_{a})=6=-\#({\bf A}_{b}) \label{NoAab}
\eeq
and the required intersection numbers $(a \circ b, a \circ b')=(3,0)$ arise provided that
\beq
x^ax^b=-y^ay^b=z^az^b=-t^at^b \label{xyz}
\eeq
Similarly,  on the {\bf (a,e,f)B} lattice, when both $a$ and $b$ are of type IV, there are solutions of the form 
\bea
A^a_p=(x^ay^a,2,-x^ay^a,-2), && (\alpha^a_i,\tilde{\alpha}^a_i)=(2y^az^a,y^at^a,-2y^az^a,-y^at^a) \\
A^b_p=(0,0,-x^by^b,2), && (\alpha^b_i,\tilde{\alpha}^b_i)=(0,0,-2y^bz^b,-y^bt^b)
\eea 
when
\bea
\frac{x^ay^a}{4}=&-{\rm Im} \ \tau _3&=\frac{x^by^b}{4} \\
X^a=&\frac{5}{4}&=X^b
\eea
These too satisfy eqs (\ref{NoAab}) and have the required intersection numbers when
\beq
x^ax^b=y^ay^b=z^az^b=-t^at^b
\eeq

Without loss of generality, we identify $a$ as the $SU(3)$ stack, and $b$ as the $SU(2)$ stack. To avoid further non-abelian gauge symmetries, all remaining stacks $\lambda$ must consist of a single $D6$-brane  with $N_{\lambda}=1$. Given the fairly limited number of solutions for $a$ and $b$, the intersection numbers $(a \circ \lambda, a \circ \lambda')$ and $(b \circ \lambda, b \circ \lambda')$ with an arbitrary (supersymmetric) stack $\lambda$ are also limited in number and highly correlated. As already noted, unavoidably we have $6q^c_L$ states arising in the antisymmetric $\bar{\bf 3}$ representation of $SU(3)$ on the stack $a$; if $y_a=1/6$ these are $6d^c_L$, whereas if $y_a=-1/3$ they are $6u^c_L$. Thus in these models the minimal quark-singlet spectrum arising from the intersections of $a$ with  other stacks $\lambda$, and their orientifold images $\lambda '$, is $3\bar{d}^c_L+3u^c_L$ when $y_a=1/6$, and $3\bar{u}^c_L+3d^c_L$ when $y_a=-1/3$. In both cases we must therefore impose the constraint $|a \circ \lambda | + |a \circ \lambda'| \leq 6$ on any one of the other stacks. The intersections of the $b$ with other stacks $\lambda$ yield doublets that must be identified  either as lepton  $L$ and Higgs $H_d$ doublets, if $Y=-1/2$, or $H_u$ doublets if $Y=1/2$.  The supersymmetric Standard Model has $3L+H_u+H_d$, so we should also impose the constraint $|b \circ \lambda| + |b \circ \lambda'| \leq 5$ on any single stack. 
 With $a$ and $b$  both of  the same type, I or IV, and on both the {\bf (a,e,f)B} and {\bf (b,c,d)B} lattices, the allowed intersection numbers, subject to the constraints described above, are displayed in Table \ref{abcmultiI}.
\begin{table}
 \begin{center}
\begin{tabular}{||c |c||} \hline \hline
$(a \circ \lambda, a \circ \lambda')$ & $(b \circ \lambda, b \circ \lambda')$  \\ \hline \hline
$(-1,-1)$&$(2,2)$\\
$(-2,-2)$& $(1,1)$\\
$(0,6)$&$(-3,0)$ \\
$(6,0)$&$(0,-3)$ \\
 \hline \hline
\end{tabular}
\end{center} 
\caption{ \label{abcmultiI} Correlations between intersection numbers of the $SU(3)$ stack $a$ and those of the $SU(2)$ stack $b$ when $(a \circ b, a \circ b')=(3,0)$.}
 \end{table}

In both cases, since the only negative intersection numbers for $a \circ \lambda$ are invariably accompanied by negative intersection numbers $a \circ \lambda '$, and {\it vice versa}, it is clear that we can never get {\em just} the required $3(\bar{\bf 3})+ 3 ({\bf 3})$ quark-singlet states. When $a$ and $b$ are both of type IV, this conclusion is true even if we do not impose the latter constraint $|b \circ \lambda| + |b \circ \lambda'| \leq 5$. However, if they are both of type I, then it can be satisfied, but only at the expense of having at least 12 doublets at the intersections of $b$ with $\lambda$ and $\lambda'$. The conclusion is that, at least within the range of parameters searched, we can{\em not} get the quark-singlet spectrum even of this Standard-like model. 

\section{Discussion}
We have investigated whether there is scope to construct supersymmetric Standard Models in type IIA intersecting-brane theories compactified on an orientifold with a $\mathbb{Z}_{12}$ point group. We focussed on the $\mathbb{Z}_{12}$-II case because, as discussed in \S \ref{Z12orb}, the $\mathbb{Z}_{12}$-I case does not have enough independent 3-cycles to make a viable model likely. The $SO(8) \times SU(2) \times SU(2)$ lattice has been used; the $F_4 \times SU(2) \times SU(2)$ case is equivalent. A bulk 3-cycle then consists of a 2-cycle on the $SO(8)$ lattice times a 1-cycle on the $SU(2)\times SU(2)$ torus $T^2_3$, and we have restricted attention to the case when the 2-cycle is factorisable in the sense discussed in \S \ref{Sfact}. It is possible to find models with the correct supersymmetric Standard Model quark-doublet content. All examples have $(a \circ b, a \circ b')=(3,0)$ or $(0,3)$ and possess 6 copies of either $d^c_L$ or $u^c_L$ quark singlets, depending on the values of $y_a$. Thus, some vector-like matter is inevitable. All examples have non-abelian gauge coupling constant unification in the sense that $\alpha _a=\alpha_b$ at the string scale, but we have not found it possible to obtain the minimal quark-singlet structure described in the previous section..


\appendix
\section{The $\mathbb{Z}_{12}$-I orientifold}

 The six independent invariant exceptional 3-cycles on the $\mathbb{Z}_{12}$-I orbifold may be chosen as follows:
\bea
\epsilon _1:=(1+\theta +\theta ^2+ ... +\theta ^{5})f_{4,4} \otimes \pi _5&=& 2[(f_{4,4}-f_{1,6}) \otimes \pi _5+(f_{1,6}-f_{4,5}) \otimes \pi _6]\\
\tilde{\epsilon} _1:=(1+\theta +\theta ^2+ ... +\theta ^{5})f_{4,4} \otimes \pi _6&=& 2[(f_{4,5}-f_{1,6}) \otimes \pi _5+(f_{4,4}-f_{4,5}) \otimes \pi _6] \\
\epsilon _2:=(1+\theta +\theta ^2+ ... +\theta ^{5})f_{4,1} \otimes \pi _5&=&(f_{4,1}-f_{6,4}+f_{4,6}-f_{5,6}) \otimes \pi _5+ \nonumber \\
 &&+(f_{6,4}-f_{6,6}+f_{5,6}-f_{5,5}) \otimes \pi _6 \\
 \tilde{\epsilon} _2:=(1+\theta +\theta ^2+ ... +\theta ^{5})f_{4,1} \otimes \pi _6&=&(-f_{6,4}+f_{6,6}-f_{5,6}+f_{5,5}) \otimes \pi _5+  \nonumber \\
&&+(f_{4,1}-f_{6,6}+f_{4,6}-f_{5,5}) \otimes \pi _6 \\
\epsilon _3:=(1+\theta +\theta ^2+ ... +\theta ^{5})f_{5,1} \otimes \pi _5&=&(f_{5,1}-f_{6,1}+f_{6,5}-f_{5,4}) \otimes \pi _5+ \nonumber \\
 &&+(f_{6,1}-f_{1,4}+f_{5,4}-f_{1,5}) \otimes \pi _6 \\
\tilde{\epsilon} _3:=(1+\theta +\theta ^2+ ... +\theta ^{5})f_{5,1} \otimes \pi _6&=&(-f_{6,1}+f_{1,4}-f_{5,4}+f_{1,5}) \otimes \pi _5+  \nonumber \\
&&+(f_{5,1}-f_{1,4}+f_{6,5}-f_{1,5}) \otimes \pi _6 
\eea
Then
\bea
\epsilon _j \circ \epsilon _k=&0&=\tilde{\epsilon} _j \circ \tilde{\epsilon} _k  \quad j,k=1,2,3\\
{\epsilon} _j \circ \tilde{\epsilon} _k&=&-12 E_j \delta _{j,k} \quad ({\rm no  \ summation})
\eea
where
\beq
E_1=2, \quad E_2=1=E_3
\eeq
assuming, as in eq. (\ref{fijkl}), that the self-intersection of a fixed point $f_{i,j}$ is $-2$.

 In this case the action of the point group generator $\theta$ is given in eq. (\ref{z121}). Then, with the central root $\alpha_2$ of the $SO(8)$ lattice parametrised as in eq. (\ref{alf2}), the remaining roots are given by
\bea
\alpha _1&=&-\sqrt{2}(e^{i\phi_1}\cos \theta  (1+\beta), e^{i\phi_2}\sin \theta  (1-\beta^{-1})) \label{alf11} \\
\alpha _3&=&-\sqrt{2}\beta^2(e^{i\phi_1}\cos \theta  , e^{i\phi_2}\sin \theta  ) \label{alf31} \\
\alpha _4&=&\sqrt{2}\beta^{-1}(e^{i\phi_1}\cos \theta , -e^{i\phi_2}\sin \theta)   \label{alf41}
\eea
With $\mathcal{R}$ acting as complex conjugation, as in eq. (\ref{Rzk}), it acts crystallographically on this lattice in the 6 orientations displayed in Table \ref{R21}.
\begin{table}
 \begin{center}
 \resizebox{\textwidth}{!}{%
\begin{tabular}{||c||c|c|c|c||c|c||} \hline \hline
{\bf Lattice}&\R$\alpha_1$&\R$\alpha _2$&\R$\alpha _3$&\R$\alpha _4$& $e^{-2i\phi_1}$& $e^{-2i\phi_2}$ \\ \hline \hline

{\bf a}&$-(\alpha _2+\alpha _4)$&$ \alpha _2$&$-(\alpha_2+\alpha_3)$& $-(\alpha_1+\alpha_2)$& $1$ & $1$ \\

{\bf b}&$\alpha _1+\alpha _2+\alpha _3$&$-\alpha _3$& $ -\alpha_2$& $-(\alpha_1+\alpha_2+\alpha_4)$& $ \beta^2$ & $\beta^2$ \\

{\bf c}&$-(\alpha_1+2\alpha_2+\alpha _3+\alpha_4)$&$ \alpha _2+\alpha_3$&$-\alpha_3$  & $\alpha_4$& $ \beta^{-2}$ & $ \beta^{-2}$ \\

{\bf d}&$\alpha _1$&$ -(\alpha _1+\alpha_2)$&  $-\alpha_4$    &    $-\alpha_3$& $ \beta$ & $- \beta$ \\

{\bf e}&$-(\alpha _2+\alpha _3+\alpha _4)$&$\alpha _4$ &$-(\alpha_1+\alpha_2+\alpha_4)$   &$\alpha_2$& $ \beta^{-1}$ & $- \beta^{-1}$ \\

{\bf f}&$\alpha _1+\alpha _2+\alpha _3+\alpha _4$&$-(\alpha_1+\alpha _2+\alpha_4)      $& $\alpha_1+\alpha _2$&            $-(\alpha_2+\alpha_3)$&$i$ & $-i$ \\ 

\hline \hline
\end{tabular}}
\end{center} 
\caption{ \label{R21}  The phases $\phi _1$ and $\phi _2$ for crystallographic action of \R \ on $\alpha_i \ (i=1,2,3,4)$; an overall sign of $\epsilon= \pm 1$ is undisplayed.}
 \end{table}
$\mathcal{R}$ acts crystallographically on the basis 1-cycles $\pi _{5,6}$ of the $SU(3)$ lattice in $T^2_3$ in 2 orientations:
\bea
{\bf A}: \quad \mathcal{R} \pi _5=\pi _5, \quad \mathcal{R} \pi _6=\pi _5-\pi _6 \\
{\bf B}: \quad \mathcal{R} \pi _5=\pi _6, \quad \mathcal{R} \pi _6=\pi _5
\eea
Then the action of $\mathcal{R}$ on the  invariant bulk 3-cycles defined in eqs (\ref{ro11}) and (\ref{ro21}) is given in Table \ref{Rrho1}.
 \begin{table}
 \begin{center}
\begin{tabular}{||c||c|c||} \hline \hline
{\bf Lattice} & $\mathcal{R}\rho _1$ & $\mathcal{R}\rho _2$ \\ \hline \hline
{\bf (a,f)A} & $\rho_1+\rho_2$ & $-\rho_2$ \\
{\bf (a,f)B} & $\rho_1$ & $-(\rho_1+\rho_2)$ \\ \hline
{\bf (b,e)A} & $-\rho_2$ & $-\rho_1$ \\
{\bf (b,e)B} & $-(\rho_1+\rho_2)$ & $\rho_2$ \\ \hline
{\bf (c,d)A} & $-\rho_1$ & $\rho_1+\rho_2$ \\
{\bf (c,d)B} & $\rho_2$ & $\rho_1$ \\ 
  \hline \hline
\end{tabular}
\end{center} 
\caption{ \label{Rrho1}  The action of \R \ on the invariant 3-cycles.}
 \end{table}
In this case, instead of eq. (\ref{z32}), we parametrise the 1-cycle on $T^2_3$ by
\beq
dz_3=e_5(n^{\kappa}_3+m^{\kappa}_3\beta^2)d\nu
\eeq
which gives
\bea
\Omega|_{\Pi^{\kappa}}&=&-2\sin2\theta_2e_5e^{i(\phi_1+\phi_2)}[(A_1^{\kappa}-A_2^{\kappa})\beta+A_2^{\kappa}\beta^{-1}]d\lambda\wedge d\mu \wedge d\nu \\
&:=& (X^{\kappa}+iY^{\kappa})d\lambda\wedge d\mu \wedge d\nu 
\eea
where now the bulk wrapping numbers are given by
\bea
A_1^{\kappa}&:=&a_1^{\kappa}n_3^{\kappa}+a_2^{\kappa}(n_3^{\kappa}+m_3^{\kappa}) \\
A_2^{\kappa}&:=&-a_1^{\kappa}m_3^{\kappa}+a_2^{\kappa}n_3^{\kappa}
\eea
with 
\bea
a^{\kappa}_1:=n^{\kappa}_{1,2}-n^{\kappa}_{1,3}-n^{\kappa}_{3,4} \\
a^{\kappa}_2:=n^{\kappa}_{1,3}-n^{\kappa}_{1,4}+n^{\kappa}_{2,4} 
\eea
The bulk brane is now given by
\beq
\Pi^{\kappa}=A_1^{\kappa}\rho_1+A_2^{\kappa}\rho_2
\eeq
The functions $X^{\kappa}$ and $Y^{\kappa}$ are as displayed in Table \ref{XY1}.
\begin{table}
 \begin{center}
\begin{tabular}{||c||c|c||} \hline \hline
{\bf Lattice} & $X^{\kappa}$ & $Y^{\kappa}$ \\ \hline\hline
{\bf (a,f)A} & $\sqrt{3}A_1^{\kappa}$&$A_1^{\kappa}-2A_2^{\kappa}$ \\
{\bf (a,f)B} &$2A_1^{\kappa}-A_2^{\kappa}$  &$-\sqrt{3}A_2^{\kappa}$ \\ \hline
{\bf (b,e)A} & $\sqrt{3}(A_1^{\kappa}-A_2^{\kappa})$&$-(A_1^{\kappa}+A_2^{\kappa})$ \\
{\bf (b,e)B} &$A_1^{\kappa}-2A_2^{\kappa}$  &$-\sqrt{3}A_1^{\kappa}$ \\ \hline
{\bf (c,d)A} & $\sqrt{3}A_2^{\kappa}$&$2A_1^{\kappa}-A_2^{\kappa}$ \\
{\bf (c,d)B} &$A_1^{\kappa}+A_2^{\kappa}$  &$\sqrt{3}(A_1^{\kappa}-A_2^{\kappa})$ \\
\hline \hline
\end{tabular}
\end{center} 
\caption{ \label{XY1}  The functions $X^{\kappa}$ and $Y^{\kappa}$. (A global positive factor of $R_5\sin 2\theta_2 $ for each entry is omitted).}
 \end{table}
Evidently, as claimed in \S \ref{Z12orb}, up to an overall scale, all supersymmetric stacks have the same ($\mathcal{R}$-invariant) bulk part.


\end{document}